\documentclass[preprint]{elsarticle}

\usepackage{amsmath}
\usepackage{hyperref}

\newcommand{\name}[1]{\textsl{\textsf{#1}}}
\newcommand{\code}[1]{``\textsl{\texttt{#1}}''}
\newdefinition{hyp}{Hypothesis}

\begin{document}

\begin{frontmatter}

\title{On the convergence of percolation probability functions to 
	cumulative distribution functions on square lattices with $(1,0)$-neighborhood}
\author{P.V. Moskalev}
\ead{moskalefff@gmail.com}
\address{Voronezh State Agricultural University, 
	1 Michurin street, Voronezh, 394087, Russia}


\begin{abstract} 
	We consider a percolation model on square lattices with sites weighted by beta-distributed random variables $S\sim \mathrm{Beta}(a,b)$ with a positive real parameters $a>0$ and $b>0$.
	Using the Monte Carlo method, we estimate the percolation probability $P_\infty$ as a relative frequency $P^*_\infty$ averaged over the target subset of sites on a square lattice.
	As a result of the comparative analysis, we formulate two empirical hypotheses: the first on the correspondence of percolation thresholds $p_c$ to $p_0$-quantiles (where $p_0=0.592746\ldots$) of random variables $S_i$ weighing sites of the square lattice with $(1,0)$-neighborhood, and the second on the convergence of statistical estimates of percolation probability functions $P^*_\infty(p)$ to cumulative distribution functions $F_{S_i}(p)$ of these variables $S_i$ for the supercritical values of the occupation probability $p\geq p_c$.
\end{abstract} 


\end{frontmatter}

\section{Introduction}
The percolation theory born of the conjugation of graph theory and probability, attracts researchers from many fields of science \cite{stauffer.1994, grimmett.1999}.
Parametric randomization, which is to some extent present in all percolation models, makes them one of the best for investigating various stochastic phenomena: polymerization processes in liquids and gels \cite{stauffer.1976, coniglio.1979}, phase transitions in solids \cite{aizenman.1987, golden.1998}, mass or charge transfers in heterogeneous systems \cite{guyon.1989, zeimetz.2002}, and many others.

Analyzing the structure of percolation models used in physics, we can divide them into lattice, continuous and potential models \cite{hoshen.1978, cohen.1978}.
The space of lattice percolation models forms subsets of vertices (sites) and edges (bonds) of a graph with a given topology and random weights.
If the random weights correspond only to the sites (or bonds), then we get a site (or bond) percolation model.
When combining these cases, if the random weights correspond to both sites and bonds, then we get a site-bond (mixed) percolation model \cite{kirkpatrick.1973, coniglio.1979, wierman.1984}.
Connected subsets of percolation lattice sites are usually called clusters whose probabilistic characteristics depend on: a) the neighborhood of the lattice sites; b) the distribution of a random variable that weights the percolation lattice.

If we consider neighborhoods of sites with unit radius, then on a square lattice we obtain two limiting cases \cite{packard.1985}.
The first limiting case is the von Neumann $1$-neighborhood or the $(1,0)$-neighborhood.
This neighborhood contains four sites, only one coordinate of which differs from the current site's coordinate by one $V_{1,0}(x,y)=\{(x+1,y)$, $(x,y+1)$, $(x-1,y)$, $(x,y-1)\}$.
The second limiting case is the Moore $1$-neighborhood or the $(1,\infty)$-neighborhood.
This neighborhood contains eight sites, at least one coordinate of which differs from the current site's coordinate by one $V_{1,\infty}(x,y)=\{(x+1,y)$, $(x+1,y+1)$, $(x,y+1)$, $(x-1,y+1)$, $(x-1,y)$, $(x-1,y-1)$, $(x,y-1)$, $(x+1,y-1)\}$.
Combining these cases leads us to the $(1,d)$-neighborhood, where $d$ is the exponent of the Minkowski distance \cite{kolmogorov.1957}:
\begin{equation}\label{eq:L_d(a,b)}
L_d(a,b) = \biggl(\sum\limits_{i=1}^{n}|a_i - b_i|^d\biggr)^{1/d},
\end{equation}
which is equivalent: 
to the Manhattan distance at $d=1$; 
to the Euclidean distance at $d=2$;
to the minimum or maximum distance as $d\to 0$ or $d\to\infty$: 
\begin{equation}\label{eq:L_0,inf(a,b)}
L_0(a,b) = \min\limits_{i=1}^{n}|a_i - b_i|, \qquad
L_\infty(a,b) = \max\limits_{i=1}^{n}|a_i - b_i|.
\end{equation}
We note that von Neumann and Moore $1$-neighborhoods are projections of unit circles onto a square lattice with the limiting values of the Minkowski exponent $d\to 0$ or $d\to\infty$.
The family of $(1,d)$-neighborhoods in site percolation models on square lattices opens us the possibilities for variation of all the basic parameters of percolation clusters by a continuous variable $0\leq d <\infty$.

Since a successive sample of a random variable that weights a finite percolation lattice can be interpreted as a segment of a random process realization \cite{gentle.2006}, then the cluster formed on such a lattice will be a sample of this realization.
In this case, the sample of clusters on the percolation lattice can be interpreted as a cross-section of the random process that generates these clusters.
The first numerical results with statistical analysis of site percolation models on 2D and 3D square lattices with $(1,d)$-neighborhoods were previously published by the author in Russian \cite{moskalev.2013, moskalev.2014, moskalev.2018}.

\section{Models and methods}

\subsection{Model description}

The object of research in this work is a site percolation model on a square lattice with $(1,0)$-neighborhood \cite{domany.1984}.
One of the problems solved for this model is the search for a site cluster, that is, an open site subset, connected to the starting site subset.
The cluster sites satisfy condition $s_{xy} < p$, which weights the sample values $s_{xy}$ of the random variable $S$ with the occupation probability $p$.
Lattice sites satisfying the weighted inequality are called open, those that are not~--- closed sites.
If both the start and target subsets of the lattice sites are included in the cluster simultaneously, we will call it the finite size approximation of the percolation cluster.

In the classical percolation model, sites on a square lattice are weighted by a standard uniformly distributed random variable $S\sim \mathrm{Unif}(0,1)$.
However, a uniform distribution with linear cumulative functions is quite rare in natural phenomena.
The nonlinear form of cumulative functions in similar phenomena is much more common \cite{ricciardi.2005, nowak.2008}.
In the proposed model, sites on a square lattice are weighted by beta-distributed random variables $S\sim \mathrm{Beta}(a,b)$ with real shape parameters $a>0$ and $b>0$.
With the correct selection of these parameters, the beta-distributed random variable allows us to obtain a sufficiently arbitrary form of the cumulative distribution function.
Note that $S\sim \mathrm{Unif}(0,1)\equiv \mathrm{Beta}(1,1)$ and this model can be considered as an extension of the classical percolation model for the class of continuous distributions defined on the interval $[0, 1]$.

\subsection{Algorithms and software}

The main algorithm used to generate realizations of site clusters is based on iterative joining of open sites from the current neighborhood of the cluster.
Iterations begin with a starting subset of sites and continue until the disappearance of open sites in the current neighborhood of the cluster.

To implement this algorithm, we used a free software environment for statistical computing and programming \name{R} \cite{r-project.org}.
Listing 1 shows the source code of the \code{ssi20b()} function that generates the site clusters on a square lattice with $(1,0)$-neighborhood and impermeable boundary conditions.

\begin{verbatim}
ssi20b <- function(x=33, p=0.592746, 
                   set=(x^2+1)/2, all=TRUE, 
                   a=1, b=1) {
  e <- c(-1, 1, -x, x)
  acc <- array(rbeta(x^2,a,b), rep(x,2))
  if (all) acc[set] <- 2
  else acc[set <- set[acc[set] < p]] <- 2
  acc[c(1,x),] <- acc[,c(1,x)] <- 1
  while (length(set) > 0) {
    acc[set <- unique(c(
      set[acc[set+e[1]] < p] + e[1],
      set[acc[set+e[2]] < p] + e[2],
      set[acc[set+e[3]] < p] + e[3],
      set[acc[set+e[4]] < p] + e[4] ))] <- 2 }
  return(acc) }
\end{verbatim}
 
This function uses the following variables to initialize:
\code{x}~--- linear dimension of a square percolation lattice; 
\code{p}~--- occupational probability for one site;
\code{set}~--- starting site subset;
\code{all}~--- if \code{all=TRUE}, then the function uses all sites from the starting subset, or else only open sites;
\code{a}, \code{b}~--- shape parameters of the beta-distributed random variable $S$, which weighs the percolation lattice sites.
This function creates the following variables at runtime:
\code{e}~--- shift vector defined by the neighborhood of the lattice site;
\code{acc}~--- array with sample values of the random variable $S$, which weighs the percolation lattice sites.

At the beginning of this listing, we specify the vector \code{e}, which determines the shifts of site indexes from $(1,0)$-neighborhood of the internal lattice site.
Next, we initialize the elements of the square matrix \code{acc} using a sample of the beta-distributed random variable $S$.
Then, all or only the open elements of the starting subset of the sites \code{set} are marked with a numeric label exceeding the largest value of the random variable $S$ that weighs the lattice sites.
To improve the performance of the function \code{ssi20b()}, all boundary lattice sites are marked as closed.
Next, we define a loop with a precondition for continuing the iteration by the presence of sites in the \code{set} vector.
Before the iterations begin, the \code{set} vector includes the indices of the starting site subset.
On subsequent iterations, this vector includes indexes of open sites from $(1,0)$-perimeter for the \code{set} vector sites at the previous iteration.

To estimate statistically stable characteristics of site clusters on a square lattice, we need data on the relative frequencies with which the lattice sites will be included in the sample of percolation clusters.
Listing 2 shows the source code of the \code{fssi20b()} function that calculates the site relative frequencies on a square percolation lattice with $(1,0)$-neighborhood and impermeable boundary conditions.

\begin{verbatim}
fssi20b <- function(n=1000, x=33, p=0.592746, 
                    set=(x^2+1)/2, all=TRUE,
                    a=1, b=1) {
  rfq <- array(0, rep(x,2))
  for (i in seq(n)) {
    rfq <- rfq + (ssi20b(x,p,set,all,a,b) > 1) }
  return(rfq/n) }
\end{verbatim}

In addition to the variables used to initialize the function \code{ssi20b()}, this function also requires the variable \code{n}, which determines the sample size.
At runtime, the function \code{fssi20b()} creates \code{rfq}~--- array of relative frequencies with which the lattice sites are included in the cluster sample.

The function \code{ssi20b()} shown above is based on the similar function \code{ssi20()}, previously published in the \name{SPSL} package \cite{spsl.cran}.
This package contains several functions for generating percolation clusters and their samples on 2D and 3D square lattices with different sizes, neighborhood types, occupation probabilities and starting site subsets, but with a standard uniform weighting distribution $S\sim \mathrm{Unif}(0,1)$. 
The basic characteristics of percolation models realized by functions from the \name{SPSL} package have been described by the author in previously published works \cite{moskalev.2013, moskalev.2014, moskalev.2018} in Russian.
The scope of these percolation models is limited by such stochastic phenomena that are satisfactorily described by uniformly distributed random variables with linear cumulative distribution functions.
However, as it was said above, the nonlinear form of cumulative distribution functions is much more common \cite{gupta.2004, taylor.2014}.
The function \code{ssi20b()} was developed for percolation modeling of just such phenomena.

\section{Results}

\subsection{Cluster generation}

\begin{figure}[h]\centering
	\includegraphics[width=0.6\textwidth]{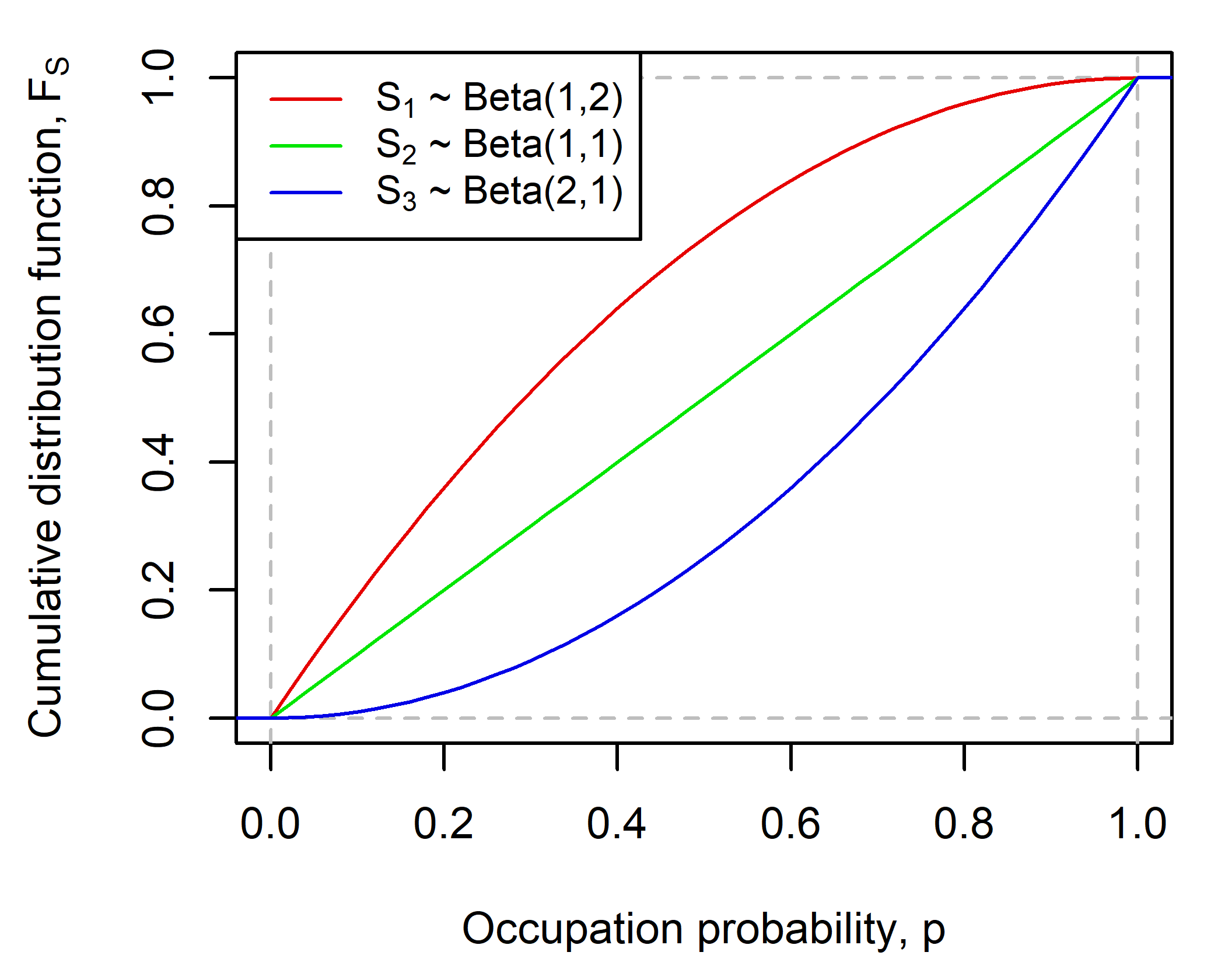}
	\caption{\small Cumulative functions $F_S(p)$ of beta-distributed weights: $S_1\sim \mathrm{Beta}(1,2)$~--- red line, $S_2\sim \mathrm{Beta}(1,1)$~--- green line, $S_3\sim \mathrm{Beta}(2,1)$~--- blue line.}
	\label{fig:weights}
\end{figure}

To generate individual clusters, we use the square lattice with $(1,0)$-neighbor\-hood, the linear size of 65 sites and three weighted distributions whose functions $F_S(p)$ are presented in Fig.~\ref{fig:weights}.
The linear cumulative function for $S_2\sim \mathrm{Beta}(1,1)$ is shown as a green line, and the nonlinear cumulative functions for $S_1\sim \mathrm{Beta}(1,2)$ and $S_3\sim \mathrm{Beta}(2,1)$ are shown as red and blue lines.

\begin{figure}\centering
	\includegraphics[width=0.45\textwidth]{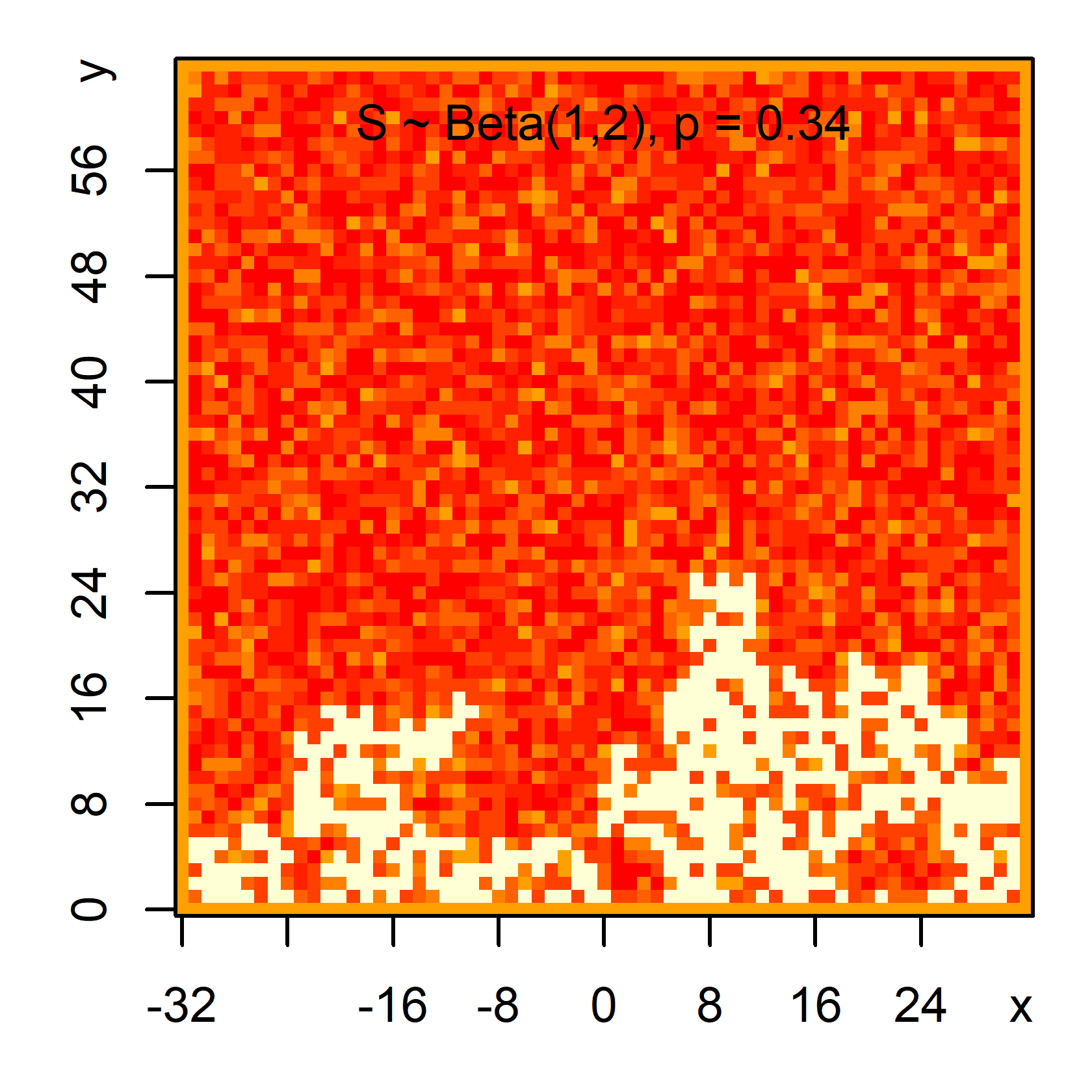}\
	\includegraphics[width=0.45\textwidth]{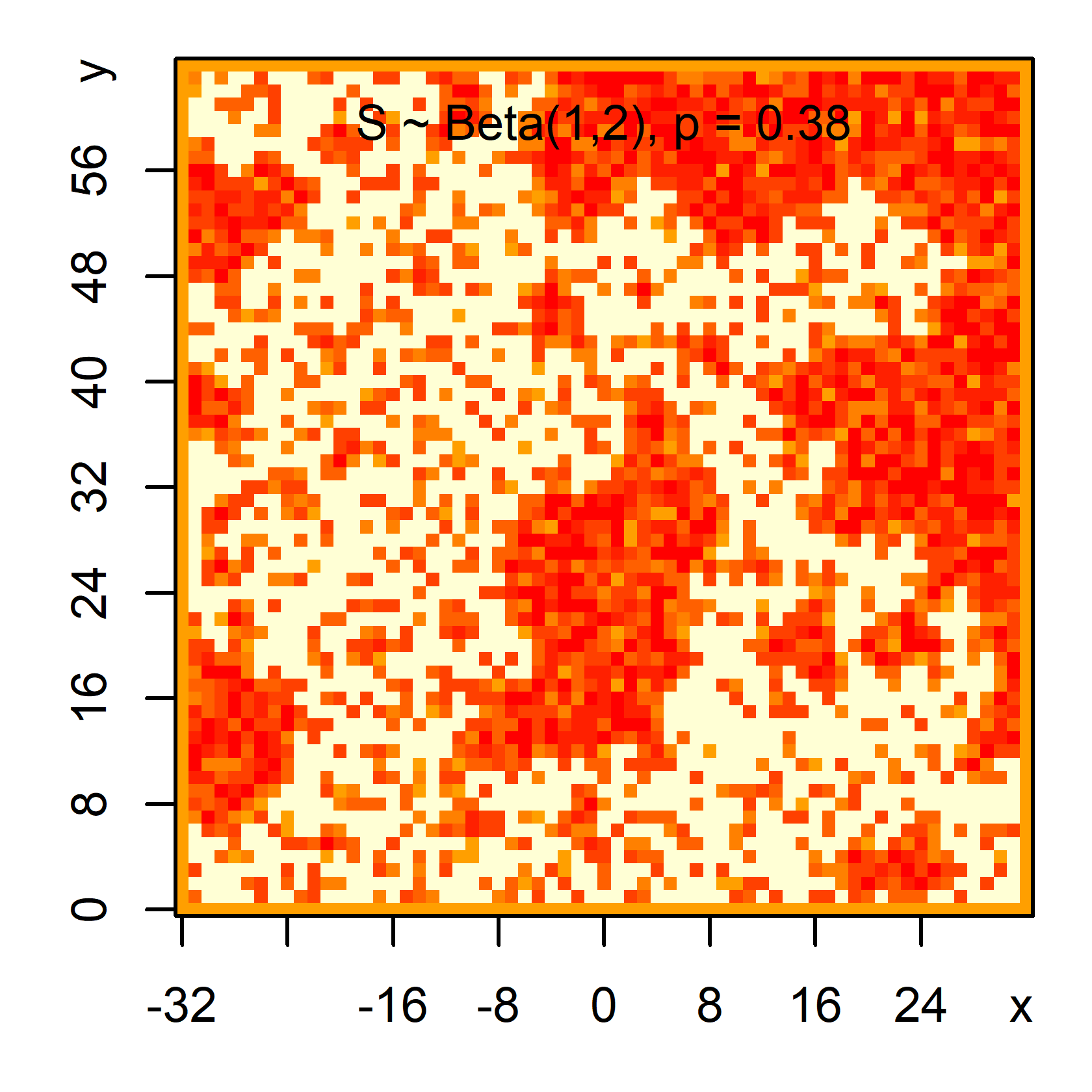}\\
	\includegraphics[width=0.45\textwidth]{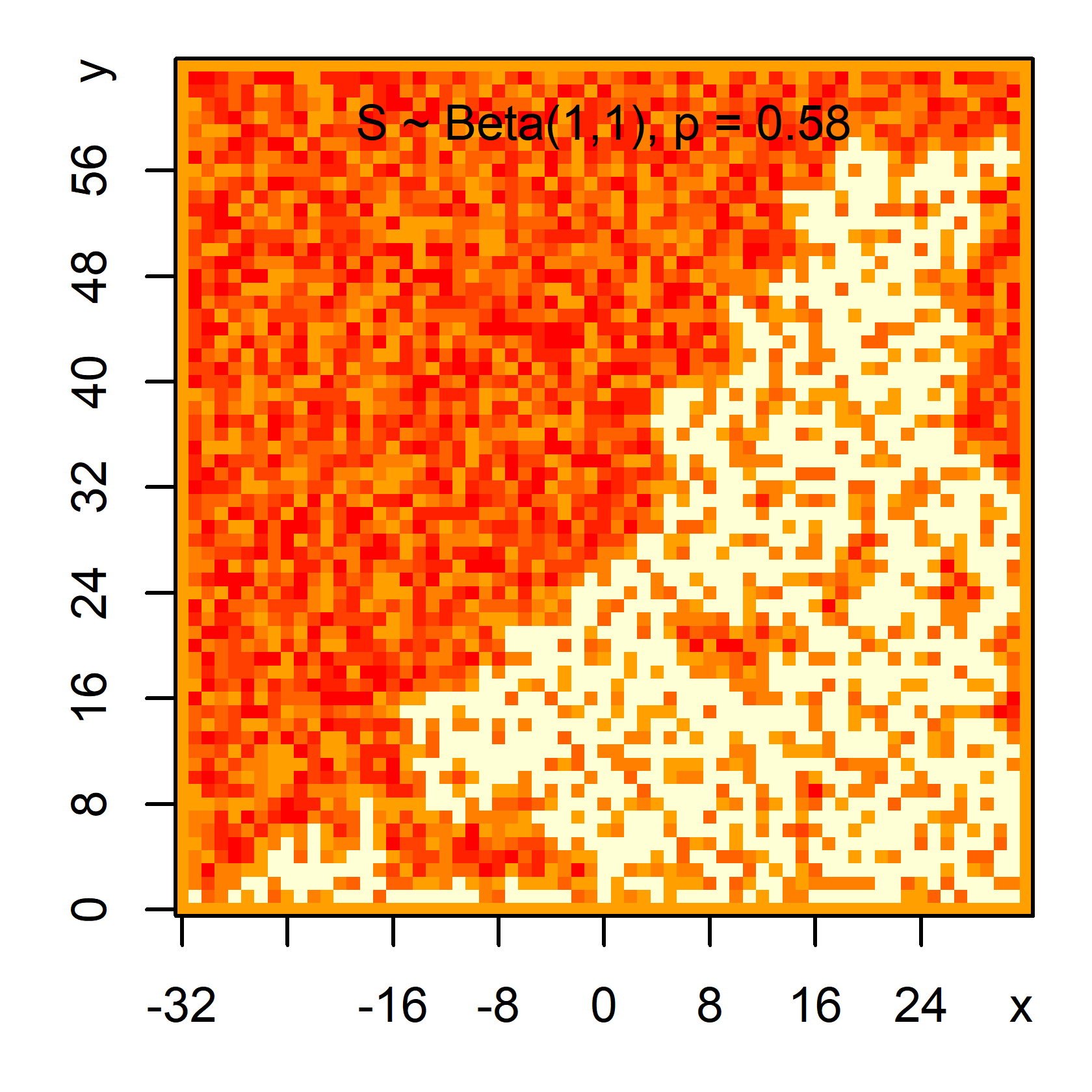}\
	\includegraphics[width=0.45\textwidth]{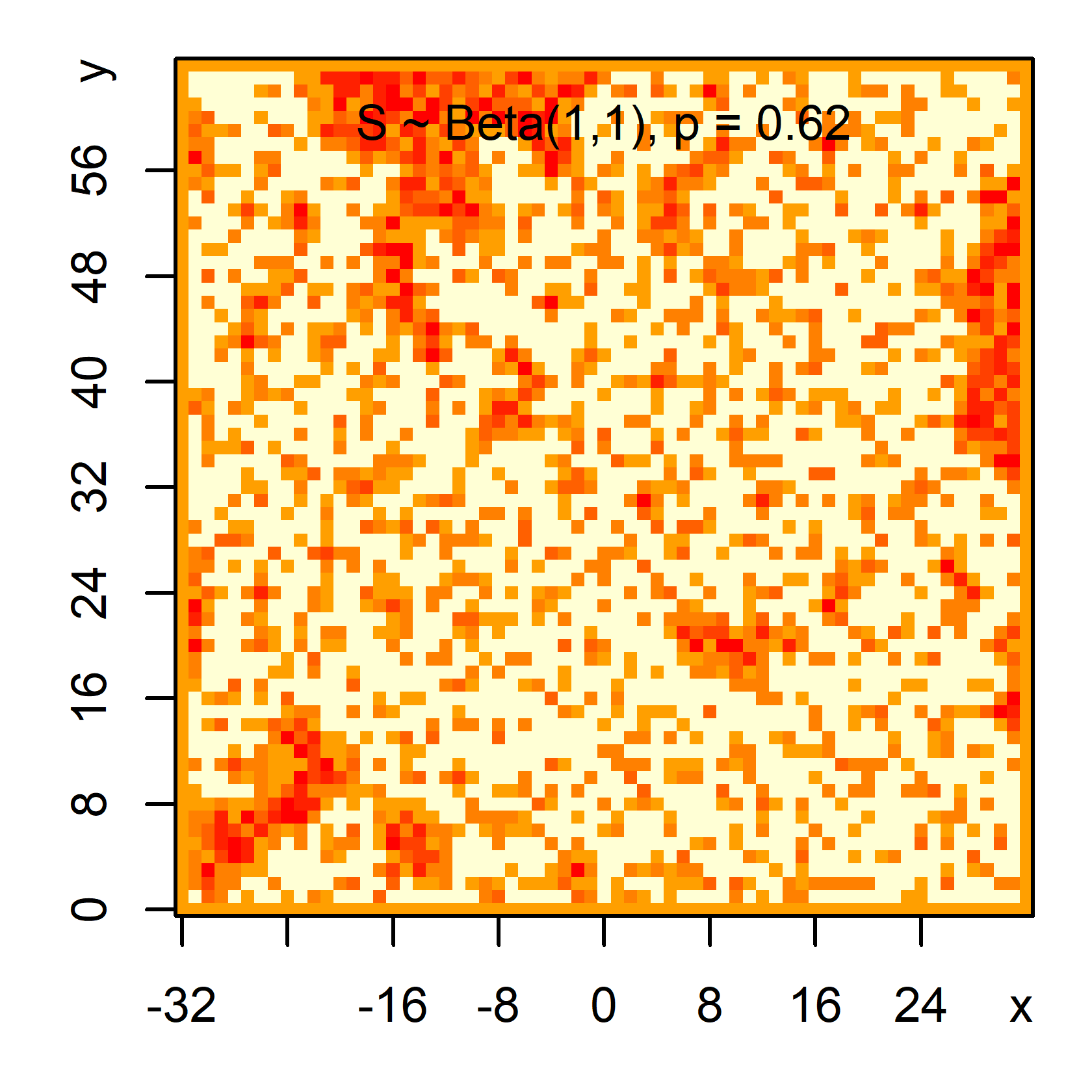}\\
	\includegraphics[width=0.45\textwidth]{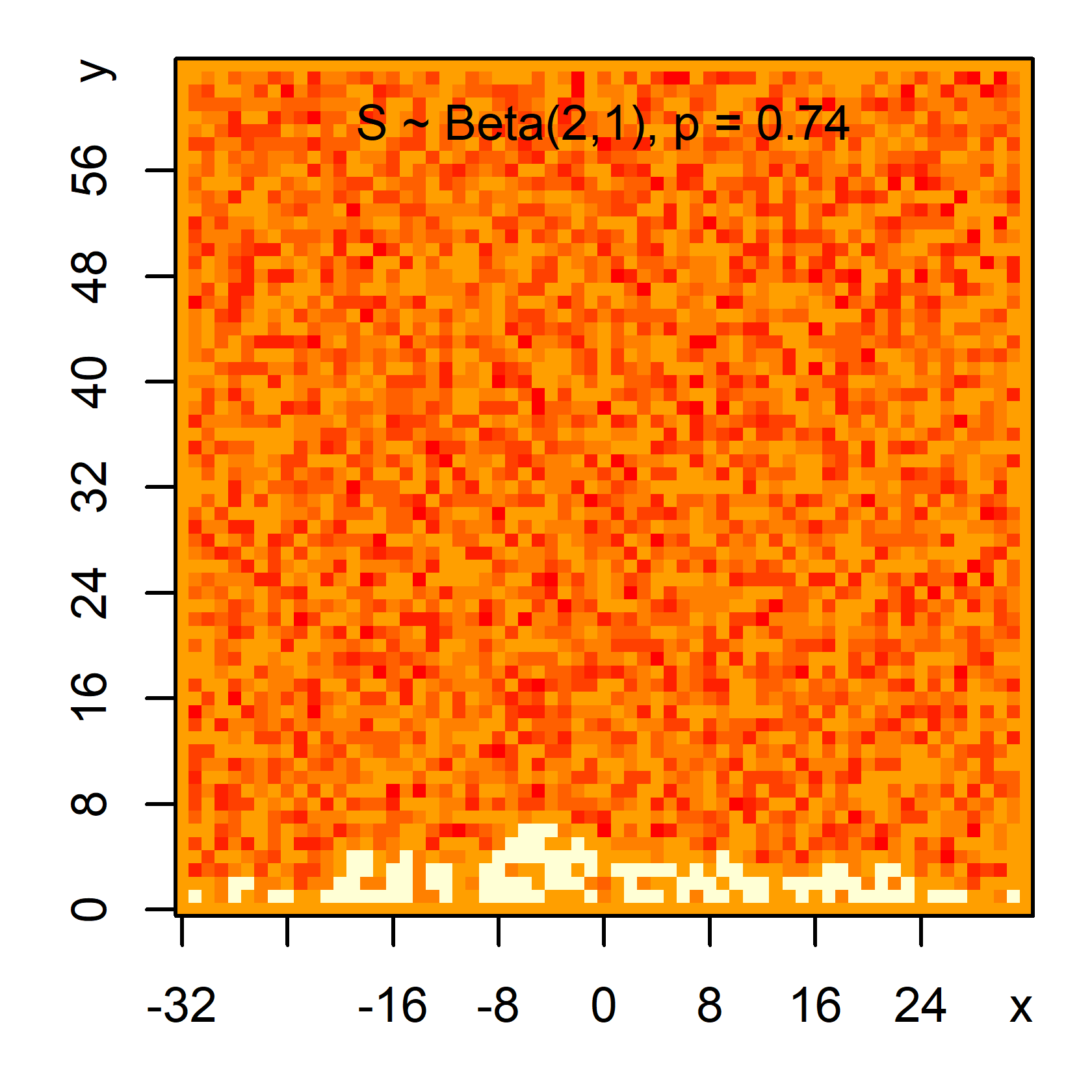}\
	\includegraphics[width=0.45\textwidth]{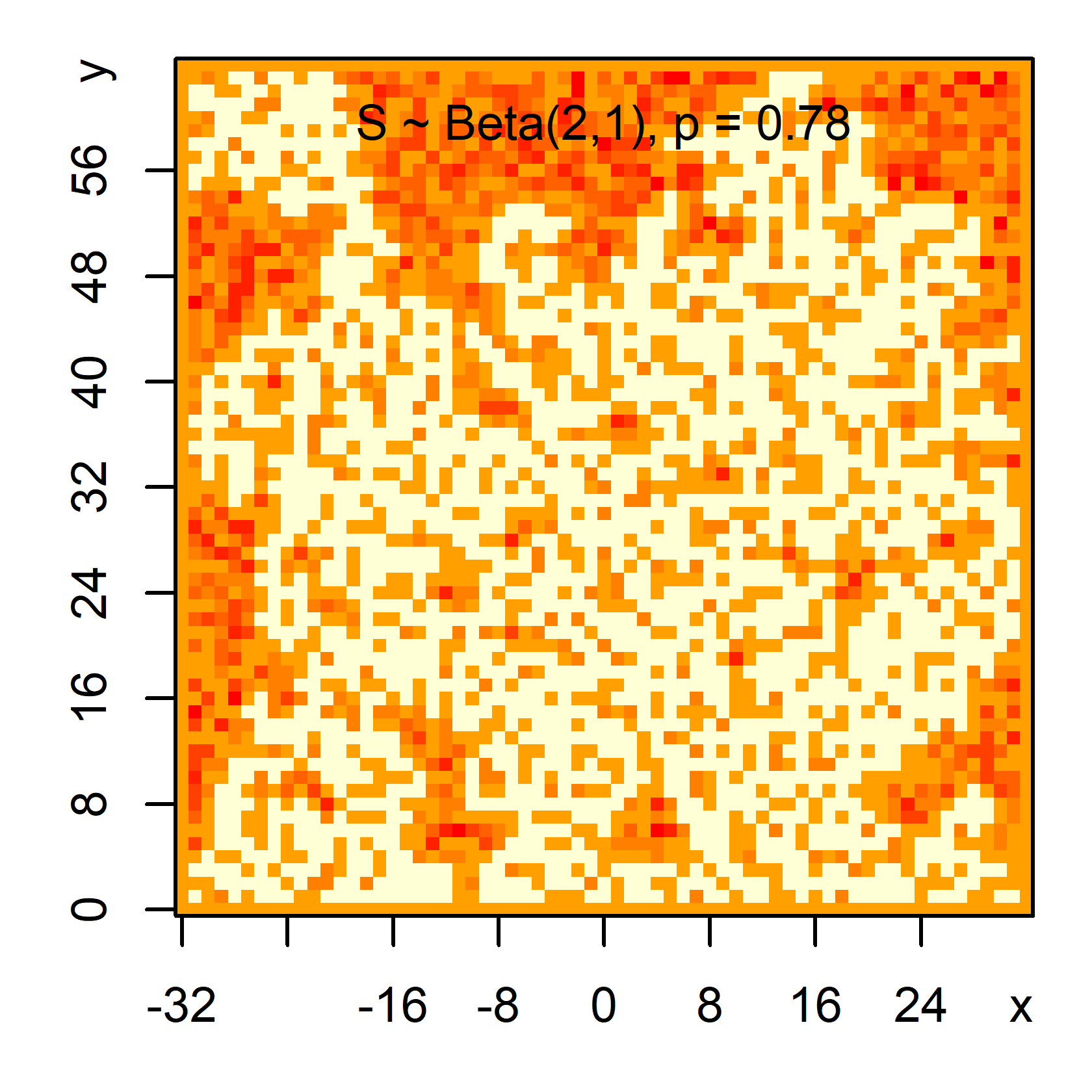}
	\caption{\small Site clusters for various weight distributions: $S_{1j}\sim\mathrm{Beta}(1,2)$~--- top row, $S_{2j}\sim\mathrm{Beta}(1,1)$~--- center row, $S_{3j}\sim\mathrm{Beta}(2,1)$~--- bottom row, and occupation probabilities: $p_{i1}<p_{c}$~--- left column, $p_{i2}>p_{c}$~--- right column.}
	\label{fig:cls}
\end{figure}

In Fig.~\ref{fig:cls} we show individual realizations of site clusters for various parameters of the percolation lattice.
As the starting subset, we selected open sites along the lower boundary of the square lattice at $y=1$.
The occupation probability in Fig.~\ref{fig:cls} changed from subcritical for $p<p_c$ in the left column, to supercritical for $p>p_c$ in the right column.
The convexity of the functions $F_S(p)$ for the weight distribution in Fig.~\ref{fig:cls} changed from negative at $S_1\sim \mathrm{Beta}(1,2)$ in the top row, through zero at $S_2\sim \mathrm{Beta}(1,1)$ in the central row, to positive at $S_3\sim \mathrm{Beta}(2,1)$ in the bottom row.

The red color in Fig.~\ref{fig:cls} corresponds to lattice sites with close to zero weights, orange to sites with weights close to one, and white to sites marked on the cluster.
In this case, the redder is the overall color of the lattice sites, the higher is the frequency of its open site.
A comparative analysis of the cluster realizations in Fig.~\ref{fig:cls} and the cumulative distribution functions $F_S(p)$ in Fig.~\ref{fig:weights} shows that the large convexity of the cumulative function leads to a shift of the percolation threshold $p_c$ to unity.

\subsection{Relative frequencies of lattice sites}

As is known, the percolation threshold $p_c$, like any other parameters calculated from individual implementations of site clusters, is the values of some random variable \cite{kesten.1982}.
To find a statistical estimate of the mathematical expectation of such a value can be found through using a sample of these values.
For this we need to perform a statistical estimation procedure for each implementation of the site cluster, and the number of these calculations increases linearly as the sample size grows.
Then, if we assume the distribution of the estimated parameter is close to normal, then to reduce the statistical error in $k$ times, we need to increase the sample size and the number of additional computations in $k^2$ times.
We can overcome this problem by moving from statistical estimation of parameters for individual implementations of site clusters to estimating the parameters of their sample as a whole.
Input data for such an estimate will be the frequency with what each of the sites on the square lattice is used in the cluster sample.

\begin{figure}\centering
	\includegraphics[width=0.45\textwidth]{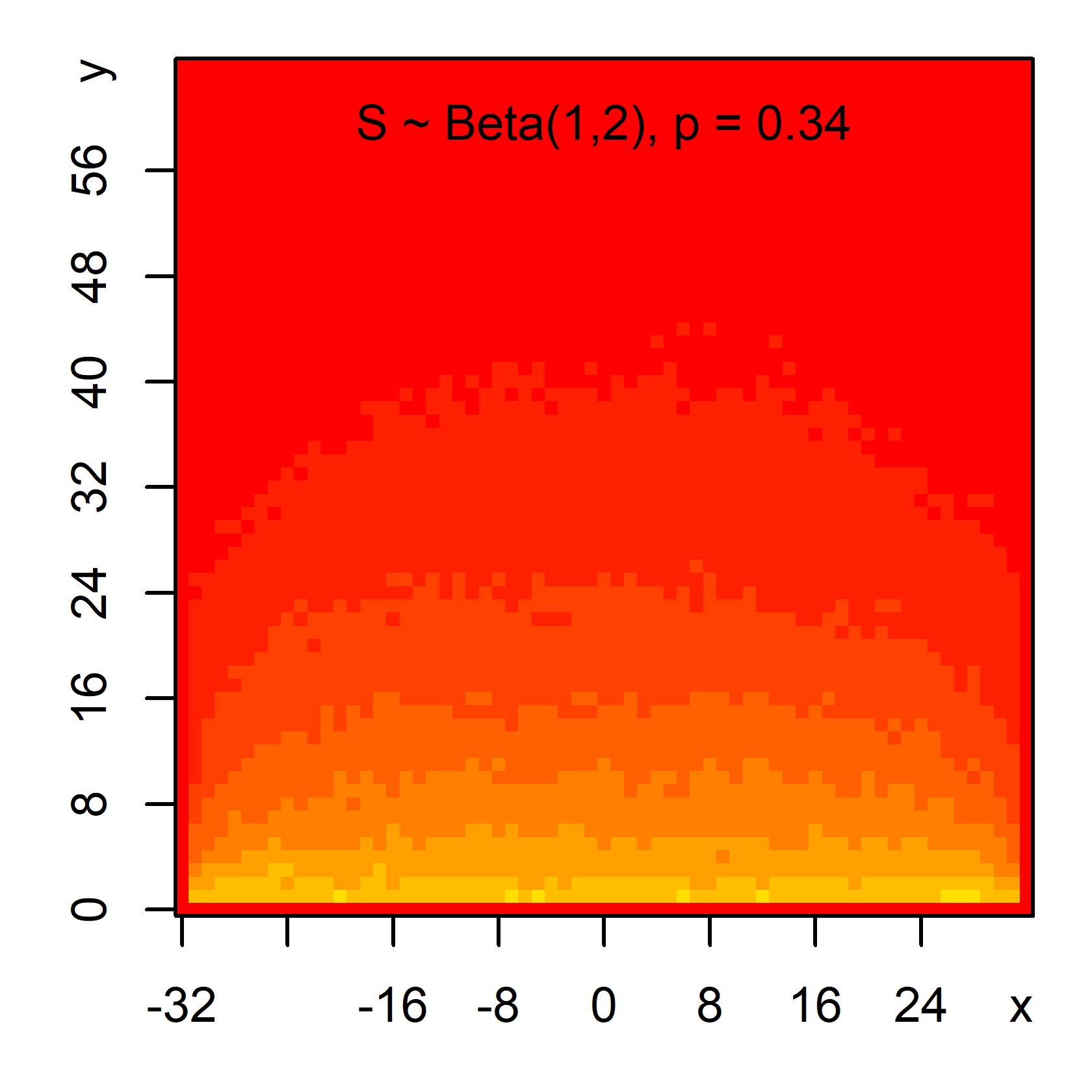}\
	\includegraphics[width=0.45\textwidth]{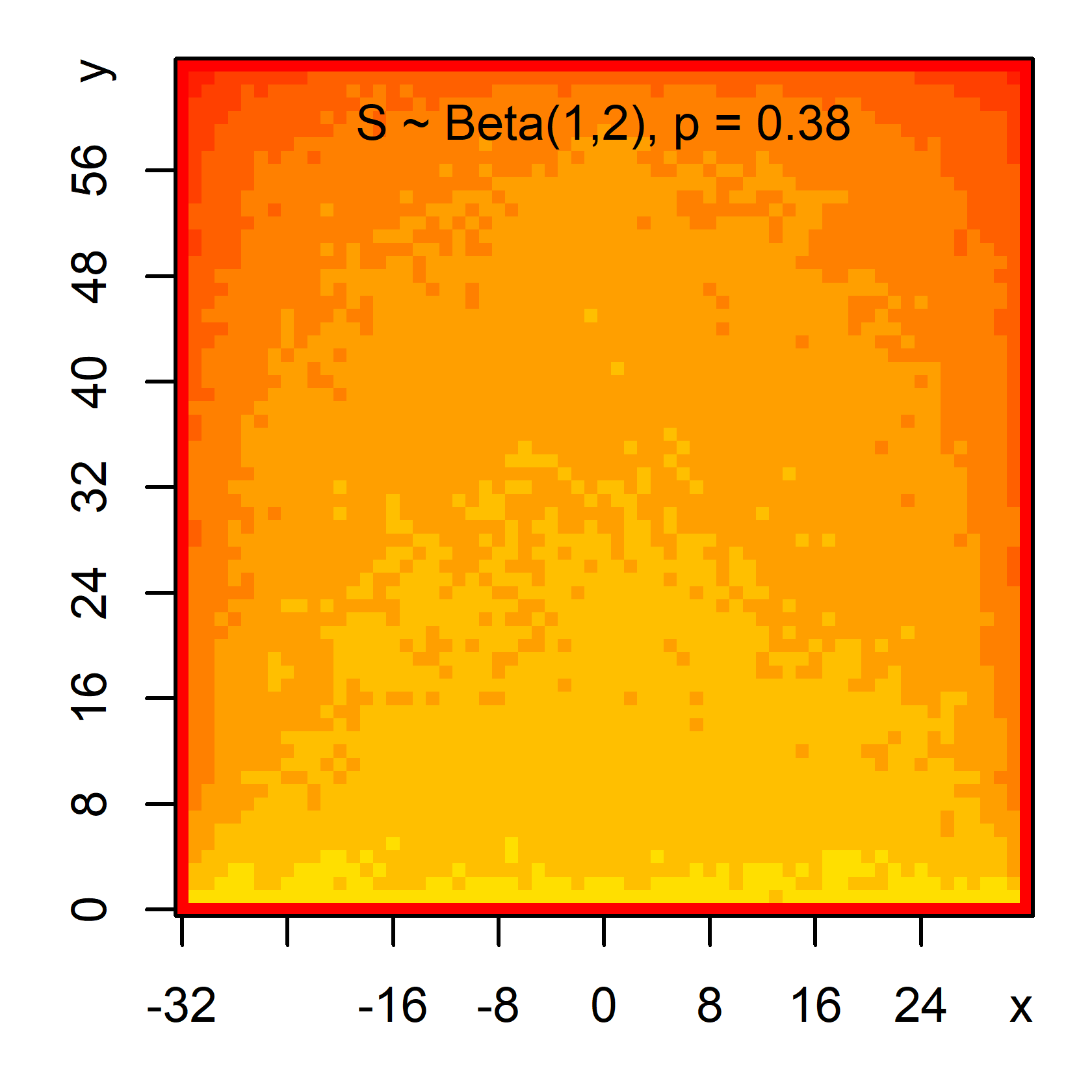}\\
	\includegraphics[width=0.45\textwidth]{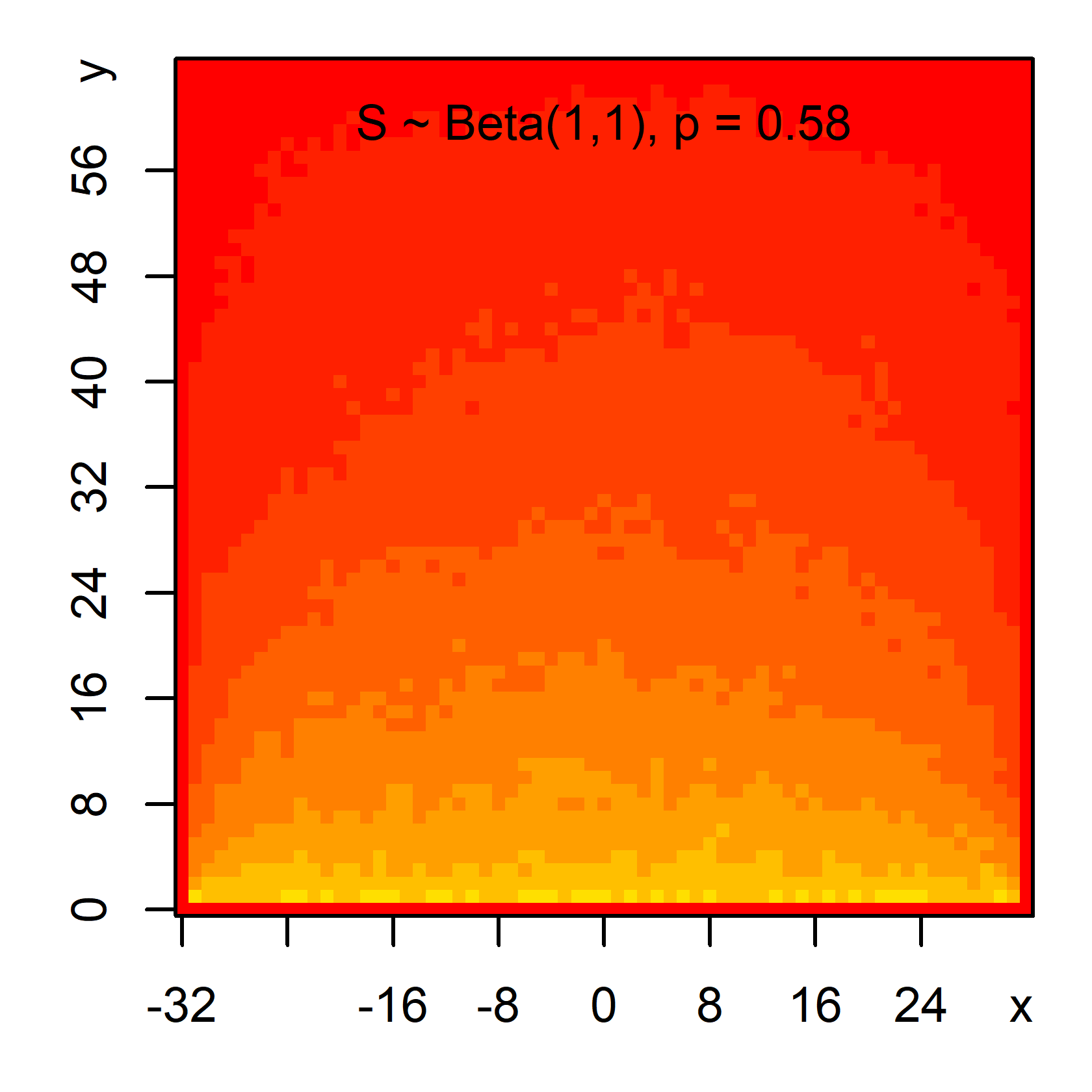}\
	\includegraphics[width=0.45\textwidth]{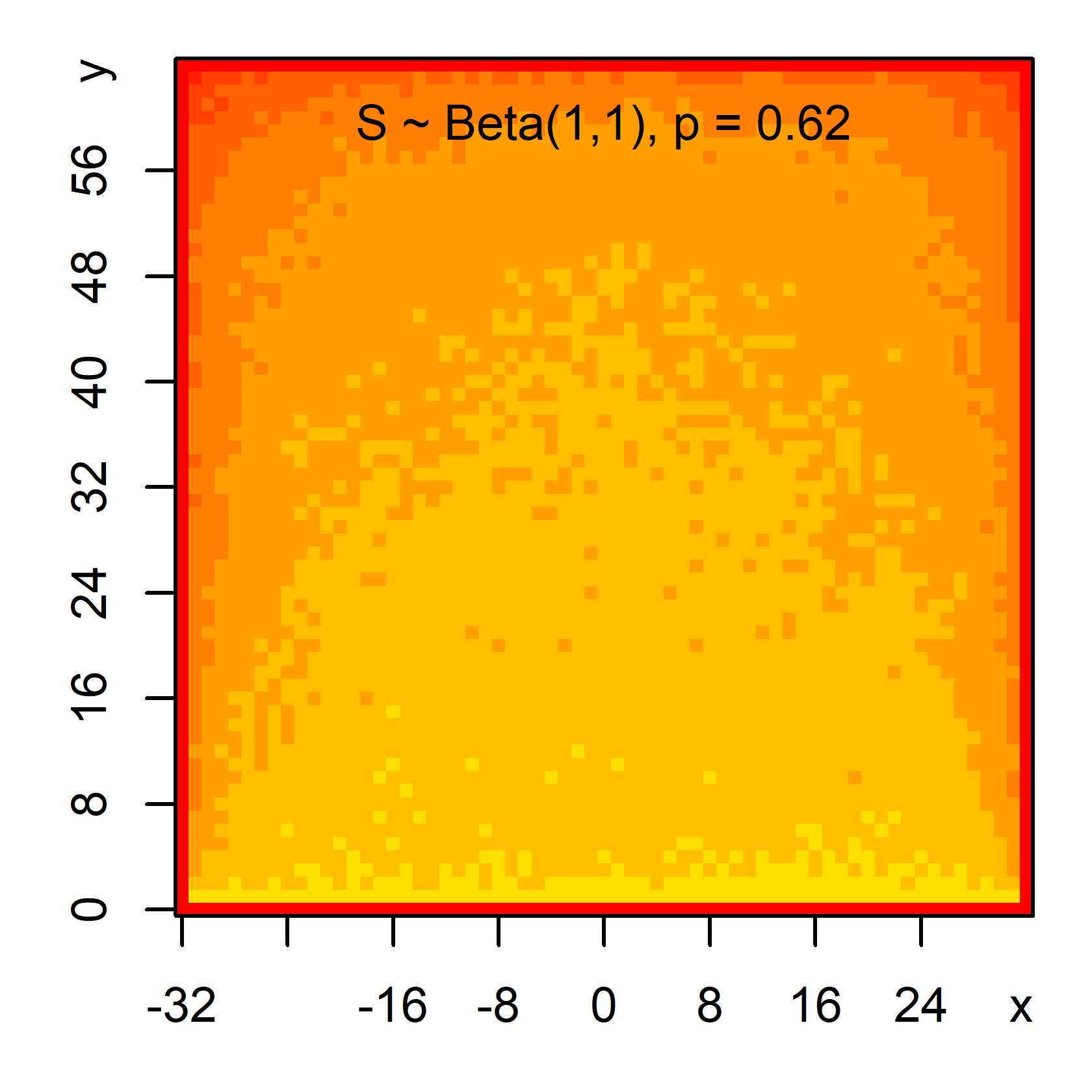}\\
	\includegraphics[width=0.45\textwidth]{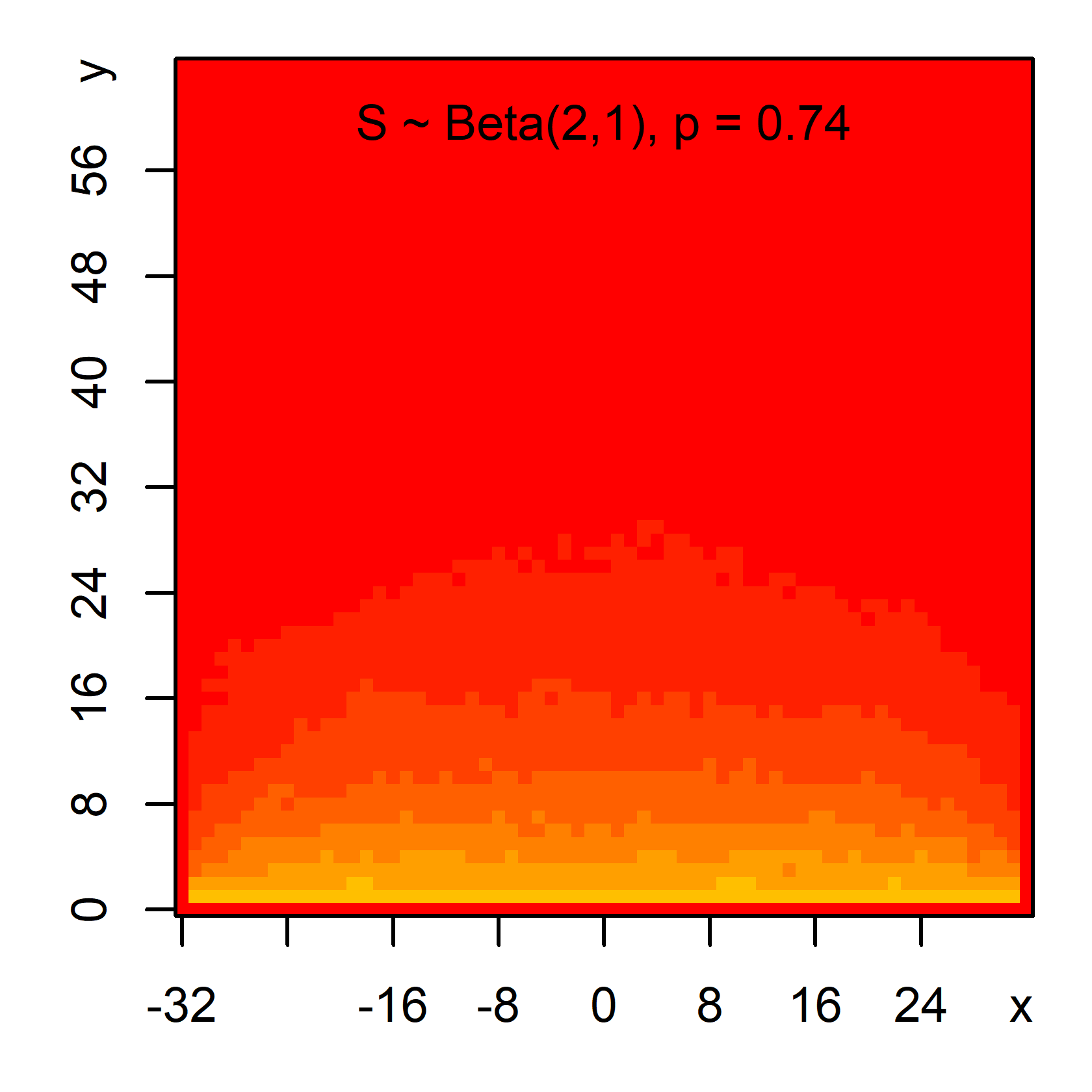}\
	\includegraphics[width=0.45\textwidth]{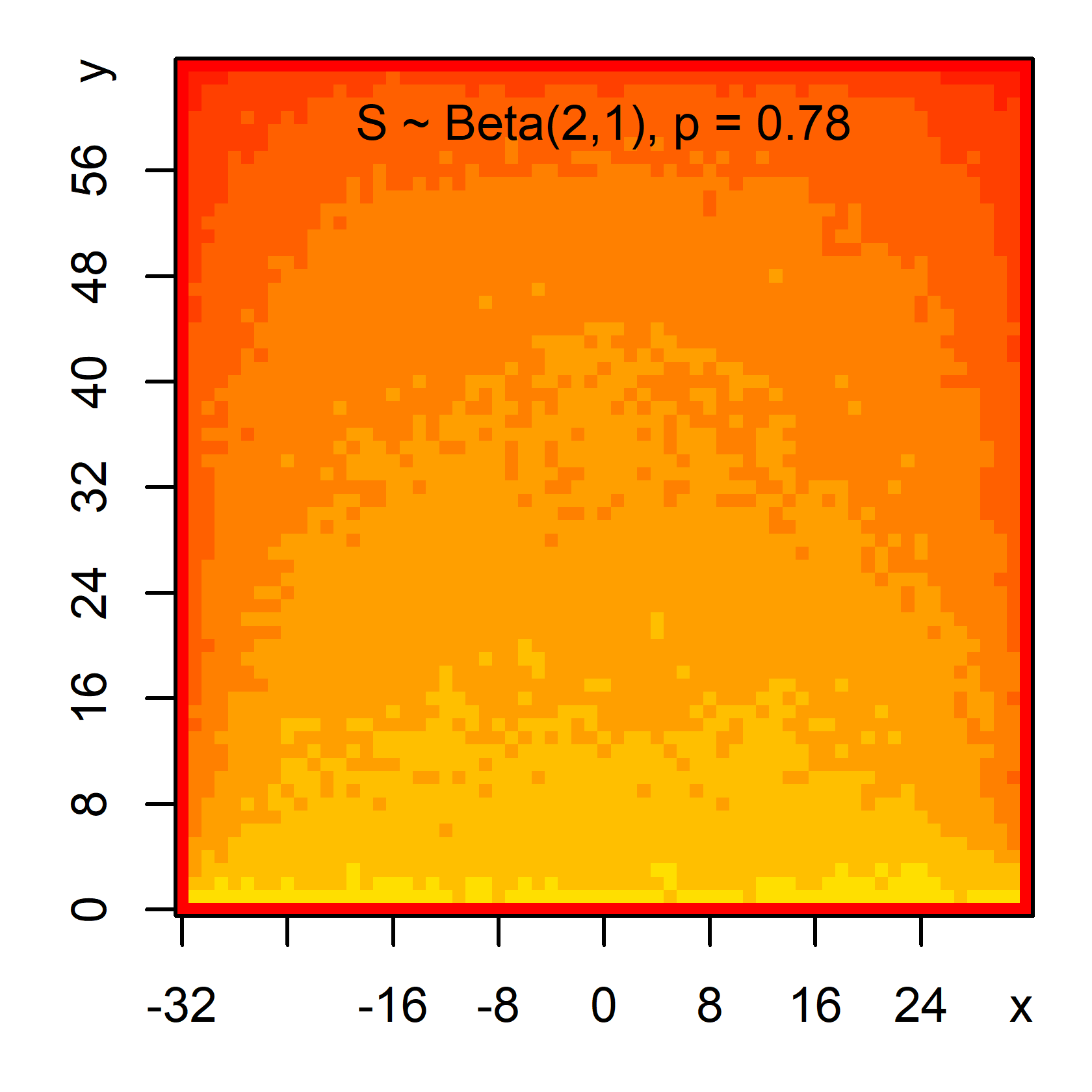}
	\caption{Relative frequencies of lattice sites for various weight distributions: $S_{1j}\sim\mathrm{Beta}(1,2)$~--- top row, $S_{2j}\sim\mathrm{Beta}(1,1)$~--- center row, $S_{3j}\sim\mathrm{Beta}(2,1)$~--- bottom row, and occupation probabilities: $p_{i1}<p_{c}$~--- left column, $p_{i2}>p_{c}$~--- right column.}
	\label{fig:rfq}
\end{figure}

In Fig.~\ref{fig:rfq} we show the relative frequencies for sites at different parameters of the percolation lattice.
The starting subset, occupational probabilities $p$ and weighting variables distributions $S$ for the percolation lattices in Fig.~\ref{fig:rfq} are identical to those shown in Fig.~\ref{fig:cls}.

The red color in Fig.~\ref{fig:rfq} corresponds to the lattice sites with relative frequencies close to zero, and the yellow color~--- to the sites with frequencies close to one.
The redder is the average color of the lattice sites, the lower is the frequency of percolation clusters.
A comparative analysis of the distributions of the relative frequencies in Fig.~\ref{fig:rfq} shows that with appropriate occupational probabilities, the spatial distribution of relative frequencies on bounded lattices will be determined by the structure of the starting and target subsets, and not by the form of the cumulative function for the weighting distribution.

\section{Discussion}

One of the main characteristics for the percolation process is the percolation cluster probability $P_\infty$.
This value corresponds to the probability of an infinite cluster on an unbounded lattice \cite{kesten.1982}, and it is defined as a function of the occupational probability $p$ at the given percolation threshold $p_c$ for this lattice:
\begin{equation}\label{eq:Pinf(p)}
P_\infty(p)\ 
\begin{cases}
\ = 0,& p < p_c;\\
\ > 0,& p\geq p_c.
\end{cases}
\end{equation}
Note that the definition \eqref{eq:Pinf(p)} is rarely used by researchers for applications, since it requires a percolation threshold $p_c$ and the form of the probability function is undefined for $p>p_c$.

For bounded lattices, the value of $P_\infty$ can be estimated from the frequency of clusters connecting the starting and target site subsets.
Using the statistics shown in Fig.~\ref{fig:rfq}, we can estimate this probability from the averaged relative frequencies of the sites along the upper boundary of the square lattice at $y = 63$ \cite{moskalev.2014}.

\begin{figure}[h]\centering
	\includegraphics[width=0.6\textwidth]{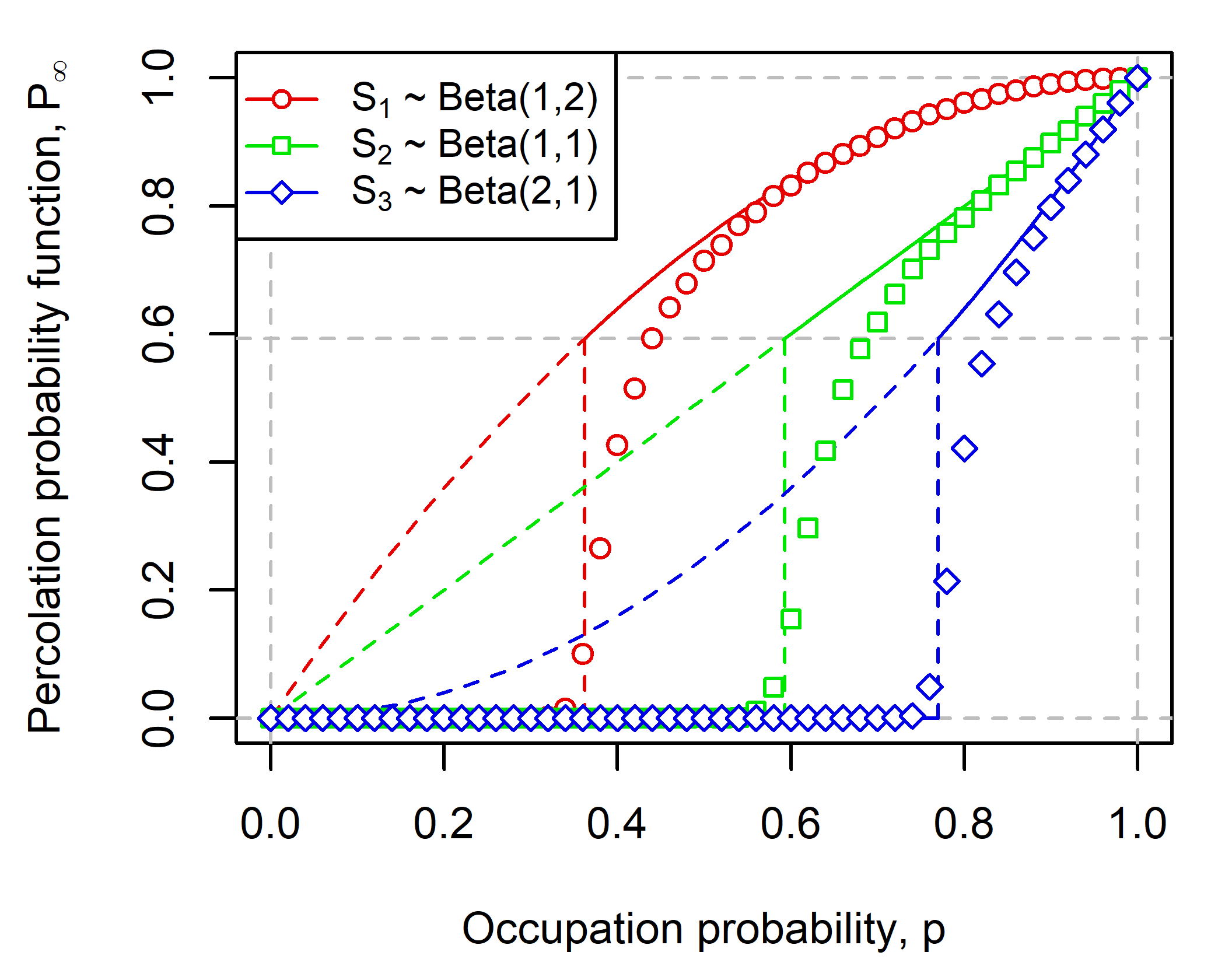}
	\caption{\small Convergence of the percolation probability function $P^*_\infty (p)$ (shown by symbols) to cumulative functions $F_S(p)$ for beta-distributed weights (shown by lines): red circles to the same color line for $S_1\sim \mathrm{Beta}(1,2)$, green squares~--- $S_2\sim \mathrm{Beta}(1,1)$, blue diamonds~--- $S_3\sim \mathrm{Beta}(2,1)$.}
	\label{fig:Pinf}
\end{figure}

Fig.~4 shows estimates of the probability function of percolation clusters on the probability of site occupation $P_\infty(p)$ for three weighted beta-distributed variables $S_1$, $S_2$, $S_3$, whose cumulative functions $F_S(p)$ were shown earlier in Fig.~\ref{fig:weights}.
Red circles, green squares and blue diamonds correspond to statistical estimates of the probability of percolation clusters $P^*_\infty(p)$ when the convexity of the weighing distribution function $F_S(p)$ changes from negative at $S_1\sim \mathrm{Beta}(1,2)$, through zero at $S_2\sim \mathrm{Beta}(1,1)$, to positive at $S_3\sim \mathrm{Beta}(2,1)$.
In Fig.~\ref{fig:Pinf} we see that the finite size estimates of the percolation probability function $P^*_\infty (p)$ describe the crossover between the subcritical clusters states for $p < p_c$ and the supercritical for $p \geq p_c$.

The vertical dashed line of green color in Fig.~\ref{fig:Pinf} corresponds to the percolation threshold $p_{c2} = 0.592746\ldots$ known from \cite{newman.2000} for a uniformly weighted square lattice.
Note that the cumulative function $F_{S_2}(p)$ of a uniformly distributed random variable $S_2\sim\mathrm{Beta}(1,1)$ coincides with its argument on $0\leq p\leq 1$.
Given this property of the cumulative function $F_{S_2}(p)$ for a uniformly distributed random variable $S_2$, the percolation threshold on a classical square lattice can be considered as a quantile with the level $p_0=0.592746\ldots$
Level $F_{S_2}(p_{c2})=p_0$ in Fig.~\ref{fig:Pinf} is shown by a horizontal dashed line of gray color.
A comparative analysis of $p_0$-quantiles for cumulative non-uniformly distribution function shown in Fig.~\ref{fig:Pinf} leads us to the first empirical hypothesis.

\begin{hyp} \itshape
	The percolation threshold $p_c$ on an unbounded square lattice with $(1,0)$-neighborhood, weighted by a continuous random variable $S$, is a priori determined by a $p_0$-quantile:
	\begin{equation}\label{eq:H1}
		p_c = F^{-1}_S(p_0),
	\end{equation} 
	where the level $p_0=0.592746\ldots$
\end{hyp}

The a priori estimates of the percolation thresholds $p_{c1}$, $p_{c2}$ and $p_{c3}$ for the weighting variables $S_1$, $S_2$ and $S_3$ found from \eqref{eq:H1} are shown in Fig.~\ref{fig:Pinf} in red, green and blue vertical dashed lines.
In Fig.~\ref{fig:Pinf} we see that all three estimates of the percolation threshold $p_{c1}=0.361835\ldots$, $p_{c2}=0.592746\ldots$ and $p_{c3}=0.769900\ldots$ are equally consistent with the results of statistical modeling.
We also see that since the percolation probability function $P_\infty (p)$ in the subcritical state converges to zero, and in the supercritical state, to the corresponding cumulative distribution function $F_S (p)$, this allows us to formulate a second empirical hypothesis.

\begin{hyp} \itshape 
	The percolation probability function $P_\infty (p)$ on an unbounded square lattice with $(1,0)$-neighborhood, weighted by a random variable $S$, has the form:
	\begin{equation}\label{eq:H2}
		P_\infty(p) = 
			\begin{cases}
				\ 0,& p < p_c;\\
				\ F_S(p),& p\geq p_c,
			\end{cases}
	\end{equation}
	where $p_c$~--- percolation threshold for this lattice; $F_S(p)$~--- cumulative distribution function for a random variable $S$.
\end{hyp}

\section{Conclusion}

In this paper we show that the relationship between such fundamental concepts as $p_0$-quantile and percolation threshold $p_c$ on a square lattice with $(1,0)$-neighborhood results from the convergence of finite-size estimates of the percolation probability function $P_\infty^* (p)$ to the cumulative distribution function $F_S(p)$ of a random variable $S$ that weights the sites of this lattice.
The theoretical analysis of the empirical hypotheses \eqref{eq:H1} and \eqref{eq:H2} formulated in this paper is a relevant and challenging problem.
For example, one of the key tasks required to analyze the hypothesis \eqref{eq:H1} is to theoretically derive the value $p_0=0.592746\ldots$ of $p_0$-quantile.
In our opinion, this value should be related to the structure of $(1,0)$-neighborhood on a square lattice.
Successful solutions to this and other problems will help researchers more clearly understand the origins of the numerous interrelationships of modern percolation theory with other areas of mathematics, physics and computer science, and will lead to the development of more accurate percolation models of natural phenomena with nonlinear cumulative distribution functions.

\end{document}